\documentstyle[preprint,epsf,aps,eqsecnum,subfigure,floats]{revtex}
\newcommand{\beq}{\begin{equation}}
\newcommand{\eeq}{\end{equation}}
\newcommand{\beqa}{\begin{eqnarray}}
\newcommand{\eeqa}{\end{eqnarray}}

\def\npb#1{Nucl.\ Phys.\ {\bf B #1}}
\def\plb#1{Phys.\ Lett.\ {\bf B #1}}
\def\prd#1{Phys.\ Rev.\ {\bf D #1}}
\def\prl#1{Phys.\ Rev.\ Lett. {\bf #1}}

\def\gsim{\ \rlap{\raise 3pt \hbox{$>$}}{\lower 3pt \hbox{$\sim$}}\ }
\def\lsim{\ \rlap{\raise 3pt \hbox{$<$}}{\lower 3pt \hbox{$\sim$}}\ }
\newcommand{\Bsgg}{\mbox{$B^{*} \to \, B\, \gamma\,\gamma\,$}}
\newcommand{\Dsgg}{\mbox{$D^{*} \to \, D\, \gamma\,\gamma\,$}}
\newcommand{\Bsg}{\mbox{$B^{*} \to \, B\, \gamma\,$}}
\newcommand{\Dsg}{\mbox{$D^{*} \to \, D\, \gamma\,$}}
\newcommand{\Dsp}{\mbox{$D^{*} \to \, D\, \pi\,$}}
\newcommand{\Bsp}{\mbox{$B^{*} \to \, B\, \pi\,$}}
\newcommand{\Bosgg}{\mbox{${B^{*}}^{0} \to \, B^0\, \gamma\,\gamma\,$}}
\newcommand{\Dosgg}{\mbox{${D^{*}}^{0}\to \, D^0\, \gamma\,\gamma\,$}}
\newcommand{\Bosg}{\mbox{${B^{*}}^{0} \to \, B^0\, \gamma\,$}}
\newcommand{\Dosg}{\mbox{${D^{*}}^{0} \to \, D^0\, \gamma\,$}}
\newcommand{\Dosp}{\mbox{${D^{*}}^{0} \to \, D^0\, \pi^0\,$}}
\newcommand{\Bpsgg}{\mbox{${B^{*}}^{+} \to \, B^+\, \gamma\,\gamma\,$}}
\newcommand{\Dpsgg}{\mbox{${D^{*}}^{+}\to \, D^+\, \gamma\,\gamma\,$}}
\newcommand{\Bpsg}{\mbox{${B^{*}}^{+} \to \, B^+\, \gamma\,$}}
\newcommand{\Dpsg}{\mbox{${D^{*}}^{+} \to \, D^+\, \gamma\,$}}

\newcommand{\Dpsp}{\mbox{${D^{*}}^{+} \to \, D^0\, \pi^+\,$}}
\newcommand{\cbsbp}{B^{*}B\pi\,}
\newcommand{\cdsdp}{D^{*}D\pi\,}

\begin{document}

%\draft

{\tighten
\preprint{\vbox{\hbox{ }}}

\title{
Two photon decays of heavy vector mesons \,\Bsgg,
\,\Dsgg\, and the possible determination
of the $g_{B^{*}(D^{*})B(D)\pi}$ and
$g_{{B^{*}}^0B^0\gamma}$ couplings.\newline
\newline
(Revised Version)}

\author{Dafne Guetta\,$^{1}$ and   Paul Singer\,$^{2}$ }

\footnotetext{\footnotesize E-mail addresses:
$1$ firenze@physics.technion.ac.il \\
\,\,\,\,\,\,\,$2$ phr26ps@physics.technion.ac.il  }

\address{ \vbox{\vskip 0.truecm}
  Department of Physics,\\
Technion- Israel Institute of Technology,\\ Haifa
32000, Israel}

\maketitle

\begin{abstract}%

We study the novel decays \Bsgg and \Dsgg using
the framework of the Heavy Meson Chiral Lagrangian
(HM$\chi$L) to leading order in chiral
perturbation theory. The branching ratios of these
decays are expressed in terms of the strong
$g_{B^{*}(D^{*})B(D)\pi}$  and the electromagnetic
$g_{B^{*}(D^*)B(D)\gamma}$ couplings, thus
providing a possible tool for their determination.
In the charm case, using the experimentally
determined ratios $({D^*}^{0,+}\rightarrow
D\pi)/({D^*}^{0,+}\rightarrow D\gamma),$ we are
able to express the branching ratio as a function
of the strong coupling only. We thus find
$1.6\times 10^{-6}<{\rm Br}(\Dosgg)<3.3\times
10^{-5}$ for $0.25<g<1,$ where $g$ is the strong
coupling of HM$\chi$L. In the beauty sector, the
${\rm Br}(\Bosgg)$ which we estimate to be in the
$10^{-7}-10^{-5}$ range is a function of both
$g_{B^{*}B\pi}$ and $g_{B^{*}B\gamma}.$ Its
behaviour does not afford an unambiguous
determination of these couplings except for the
region of high $g$ values like $g>0.6.$ The
expected two-photon differential distributions are
presented for both \Bosgg and \Dosgg, for
different values of the couplings involved.

\end{abstract}

} % end tighten

\bigskip
\leftline{ PACS number(s):  12.39.Fe, 12.39.Hg, 13.20.He}

\newpage

\section{Introduction}

The heavy vector mesons  $B^{*}$ and $D^{*}$ (of
spin-parity $1^-$)  decay via spin-flip
electromagnetic or  strong interactions to the
well studied pseudoscalar ground states $B$ and
$D.$ The decays of $D^{*}$  are known to proceed
either as a strong transition \Dsp with a final
pion with momentum of about $40\, {\rm MeV}$ or as
an electromagnetic one \Dsg, with a final photon
with momentum of about $140 {\rm MeV}.$ The
situation is different for $B^{*}$ which has a
mass of $5324.9 \pm 1.8 $ MeV; since the mass
difference $M_{B^{*}}-M_B$ is only $45.8$ MeV,
there is no strong $B^{*}$ decay and the radiative
process \Bsg is the dominant decay mode for
$B^{*}.$

In the present paper we study another possible
electromagnetic decay, the two-photon decay
processes \Bsgg and \Dsgg which were not
considered previously in the literature. In
addition to the intrinsic interest in these novel
modes, we point out that their study could provide
information on the strong couplings $g_{\cbsbp}$
and $g_{\cdsdp}$ and on the electromagnetic ones
$g_{B^*B\gamma},g_{D^*D\gamma}.$ The strong
couplings are directly related to the basic strong
coupling $g$ of the effective heavy meson chiral
Lagrangian, which describes the interactions of
heavy mesons with low-momentum pions.

There is a major dissimilarity between the
possibility of measuring the couplings in the
charm and beauty sectors, as a result of the
different mass difference between the respective
vector and pseudoscalar mesons. In the charm
sector, the experimentally measured branching
ratios of the ${D^*}^0,{D^*}^+$ decays into the
allowed $D\pi$ and $D\gamma$ modes lead to
relations between $g_{D^{*}D\pi}$ and
$g_{D^{*}D\gamma}.$ Henceforth, the \Dsgg decay
under study here is expressible in terms of the
strong coupling only and would provide a
convenient tool for its measurement.

On the other hand, in the beauty sector where the
\Bsp decay is forbidden by phase space, the $g_{B^{*}B\pi}$
coupling is not directly accessible. And although
the \Bsg decay is experimentally detected, the
direct measurement of its strength is an unlikely
proposition at present, in view of the smallness
of the expected value of its decay width.

However, as we show in this paper, the two photon
decay \Bsgg branching ratio turns out to be a
function of $g_{B^{*}B\pi}$ and
$g_{B^{*}B\gamma},$ which opens the possibility for
their determination, especially in particularly
favorable regions of the $[g_{B^{*}B\pi},
g_{B^{*}B\gamma}]$ parameter space.

Our analysis singles out the neutral modes
\mbox{${B^{*}}^{0} ({D^{*}}^{0})\to \,
B^0 (D^0)\, \gamma\,\gamma\,$} as the more relevant
ones   in relation with the  determination of the
the couplings under consideration. 
 In addition to the
direct radiative transition which is the sole
contribution in neutral decays, and on which we
concentrate in this paper, there is also the
two-photon decay arising from bremsstrahlung
in the charged
\mbox{${B^{*}}^{+} ({D^{*}}^{+})\to \,
B^+ (D^+)\, \gamma\,$} channel. Since this
radiation overwhelms the direct mode, as we will
show, one has to resort to the investigation of
the neutral modes if one aims for a cleaner
determination of the strong couplings.

In section II we review the present status of the
main decays  of $B^{*}$ and $D^{*},$ with which
the rare two photon decays must be compared. In
section III we present the theoretical framework
of our approach.  Section IV contains the explicit
treatment of the decay amplitudes. In the
last section we summarize our predictions and we
discuss certain features of the calculation.

\section{The experimental and theoretical status
of $B^{*}$ and $D^{*}$  principal decays}

The vector mesons $B^{*}$ were firstly observed
\cite{cusb}  by the CUSB collaboration at the
Cornell Electron  Storage Ring (CESR) by detecting
the photon signal from  the radiative decay
\Bsg.
This signal of $46$ MeV photons was confirmed in
improved CUSB-II measurements, with the vector
mesons produced at CESR  at the $\Upsilon(5S)$
resonance \cite{cusb2}. Recently the process
\Bsg has been observed also at LEP with the various
detectors \cite{opal} in a sample of
over 4 milion hadronic $Z^{0}$ decays.
The rate of the $B^{*}$-meson production relative
to that of $B$ mesons is found generally to be consistent
 with the expectation
from spin counting. It is expected that this
production rate will be maintained in future $B$
experiments at hadron machines, like BTeV  at
Fermilab and LHC at CERN, where samples of the
order of $10^{10}$ $B'$s are expected. Then, a fairly
high sensitivity can be achieved in the study of
$B^{*}$ decays and measurements of $B^{*}$
branching ratios of the order $10^{-7} - 10^{-8}$
could be accessible.

In view of the small mass difference $\Delta
M(B^{*}-B)=45.78\pm 0.35$ MeV \cite{PDG}, which
forbids strong $B^{*}$ decays,the electromagnetic
transition \Bsg appears as the main decay of
$B^{*}.$ This decay has been studied in a variety
of theoretical models, including quark models
\cite{Eichten,Godfrey,Ivanov}, the chiral bag
model \cite{Singer} followed by effective chiral
Lagrangian approaches for heavy and light mesons
\cite{Cho,Amundson,Colangelo},
potential models \cite{Bank,DeFazio} and QCD sum
rules \cite{Dosch,Aliev}. The predictions of these
calculations span a range of nearly one order of
magnitude for the expected decay widths, between
$\Gamma({B^{*}}^0 ({B^{*}}^+)\rightarrow B^0(B^+)
\gamma)=0.04(0.10)$ KeV \cite{Dosch} and
$\Gamma({B^{*}}^0
({B^{*}}^+)\rightarrow B^0(B^+)
\gamma)=0.28(0.62)$ KeV \cite{Singer},
with most of the calculations
\cite{Godfrey,Ivanov,Cho,Aliev} giving values
closer to the larger values of Ref. \cite{Singer}.

The $D^{*}$ meson was discovered more than twenty
years ago \cite{Peruzzi}  and has been
subsequently studied in several experiments at
different accelerators e.g.
\cite{JADE,HRS,CLEO,ARGUS,CLEO1,ACCMOR}. The
${D^{*}}^+$ ($ M=2010.0\pm 0.5 \,{\rm MeV}$) and
${D^{*}}^0$ ($ M=2006.7\pm 0.5 \,{\rm MeV}$) have
relatively  little phase space for strong decay
into $ D + \pi.$ The current PDG averages
\cite{PDG} for the measured branching ratios of
the observed decays are  ${\rm Br}({D^{*}}^+
\rightarrow D^+\pi^0):{\rm Br}({D^{*}}^+
\rightarrow D^0\pi^+):{\rm Br}({D^{*}}^+
\rightarrow D^+\gamma)=(30.6\pm2.5)\%:(68.3\pm1.4)\%:
(1.1+2.1-0.7)\% $ and ${\rm Br}({D^{*}}^0
\rightarrow D^0\pi^0):{\rm Br}({D^{*}}^0
\rightarrow D^0\gamma)=(61.9\pm2.9)\%:(38.1\pm2.9)\%.$
The most recent experiment on $ {D^{*}}^+$ decays
\cite{CLEO1} gives more accurate branching ratios
for the three decay channels as follow
$(30.7\pm0.7)\%:(67.6\pm0.9)\%: (1.7\pm 0.6)\% .$

Although there are, by now, good data on the
branching ratios, there is still no absolute
measurement of any of the partial decay widths.
The tightest upper limit has been established by
the ACCMOR Collaboration at CERN \cite{ACCMOR}
from the measurement of 127 ${D^{*}}^+$ events
using a high-resolution silicon vertex detector,
to be $\Gamma({D^{*}}^+)<131$ KeV. The other
closest limit, obtained by the HRS collaboration
\cite{HRS}, gives upper limits of 1.1 MeV and 2.1
MeV for the total decay widths of the charged and
neutral ${D^{*}}'$ s.

The decays of ${D^{*}}$ have also been treated
extensively in a plethora of theoretical models.
Many of the papers we mentioned concerning the
${B^{*}}$ decay
\cite{Eichten,Godfrey,Ivanov,Singer,Cho,Amundson,Colangelo,Bank,DeFazio,Dosch,Aliev}
discuss also the ${D^{*}}$ decays. In addition, we
want to mention the early approaches
\cite{Thens,Pham} with effective Lagrangian
including symmetry-breaking, a  relativistic quark
model \cite{Donnel}, the study of ${D^{*}}$ decays
using the chiral-bag model which contains pion
exchange effects (pion loops) \cite{Miller}, the
use of QCD sum rules \cite{Belyaev}, a chiral
model with $ M_{c}\rightarrow \infty $
\cite{Troytskaya} and the comprehensive analysis
of Kamal and Xu \cite{Kamal}. Recently
\cite{Stewart}, the strength of the various decay
channels of $D^*$ has been extracted from an
analysis of the experimental branching ratios by
the use of the chiral perturbation theory.

In the ${D^{*}} $ case, the theoretical
calculations again span an order of magnitude
range for the prediction of the absolute decay
widths, from a small width of
$\Gamma({D^{*}}^0)\simeq (3-10)$ KeV
\cite{Dosch,Troytskaya} to $\Gamma({D^{*}}^0)\simeq
(60-120)$ KeV \cite{Cho,Pham,Donnel},
 including fairly  large uncertainties. It should be
emphasized at this point that the chiral bag
calculation ($\chi$) has offered the best
estimation for  the branching
ratios \cite{Miller} ${\rm Br}^{\chi}({D^{*}}^+
\rightarrow D^+\pi^0):{\rm Br}^{\chi}({D^{*}}^+
\rightarrow D^0\pi^+):{\rm Br}^{\chi}({D^{*}}^+
\rightarrow D^+\gamma)=31.2\%:67.5\%:1.3\%$
and ${\rm Br}^{\chi}({D^{*}}^0
\rightarrow D^0\pi^0):{\rm Br}^{\chi}({D^{*}}^0
\rightarrow D^0\gamma)=64.3\%:35.7\%.$

The recent experiments have confirmed
\cite{PDG,CLEO1} these relative ratios and
dispersed the puzzling features which prevailed
previously concerning the radiative branching
ratio and the relative ratios of ${D^{*}}^+$ strong channels
(see, e.g. Ref.\cite{Kamal}). The prediction of
the chiral bag model \cite{Miller} is
$\Gamma({D^{*}}^+\rightarrow all) =79$ KeV,
$\Gamma({D^{*}}^0\rightarrow all)= 59.4$ KeV.
Several of the other calculations result in fairly
similar values \cite{Eichten,Godfrey,Kamal} as
well as predicting for $\Gamma({D^{*}}^0)$ a
value approximately $25\%$ smaller than for
$\Gamma({D^{*}}^+).$ There are also calculations
in which these widths are nearly equal
\cite{Colangelo,Dosch} or, on the contrary,
calculations giving
 $\Gamma({D^{*}}^+)$ to be at least twice
larger than $\Gamma({D^{*}}^0)$
\cite{Aliev,Belyaev,Troytskaya}.

The experimental and theoretical survey we
presented here is obviously of direct relevance to
our calculation as the absolute values of $B^{*}$
and $D^{*}$ widths will affect the probability of
the \Bsgg and \Dsgg future detection.

\section{A Model for Two-Photon Transition.}

Substantial progress has been made in recent years
in the treatment of the interactions of heavy
mesons containing a single heavy quark with low
momentum pions, by the use of an effective
Lagrangian \cite{Wise,Burdmann,Yan}, the so-called
``Heavy Meson Chiral Lagrangian" (HM$\chi$
L),which embodies two principal symmetries of
Quantum Chromodynamics (for a comprehensive review
of this theoretical framework and its
applications, see \cite{Casalbuoni}). At the
leading order in an $1/M_{H}$ expansion ($M_{H}$
is the mass of the heavy meson) and the chiral
limit for the light quarks ($m_{l}\rightarrow
0,l=u,d,s$), the Lagrangian carries flavor and
spin symmetry in the heavy meson sector, as well
as $SU(3)_L\otimes SU(3)_R$ chiral invariance in
the light meson one. We adopt this framework for
the calculation of the processes we study here,
namely the $B^*(D^*)\rightarrow B(D) \gamma \gamma
$ decays, and we shall use it to display the
possible usefulness  of these transitions for the
determination of the  $g_{\cbsbp}, g_{\cdsdp}$ and
$g_{B^{*}B\gamma}$ couplings.

The heavy vector ($B^*$ or $D^*$) and pseudoscalar
($B$ or $D$) mesons are represented by a $4\times
4$ Dirac matrix $H$, with one spinor index for the
heavy quark and the second one for the light
degree of freedom.
\beq
H=\frac{1+\not v}{2}\left[B^*_\mu \gamma^\mu -
B\gamma_5\right],\,\,\,\bar{H} =\gamma_0 H^\dagger
\gamma_0,
\eeq

where $v$ is the meson velocity, $v^\mu B^*_\mu
=0$ and $B_\mu^*,B$ are the respective annihilation
operators of the meson fields. We shall usually
refer henceforth to \Bsgg with the understanding
that the same treatment holds for \Dsgg. However
we shall specify the two channels separately when
numerical or other specific  features make it
necessary.

The relevant interaction term of the ${\rm
HM}\chi{\rm L},$ representing the coupling of
heavy mesons to an odd number of pions, is given
by \cite{Wise,Burdmann}
\beq
\label{lchir}
{\cal L}^{\rm int}_{{\rm HM}\chi L} = g {\rm Tr}
\left( \bar{H}_a \gamma_\mu \gamma_5 {\cal A}^\mu_{ab}H_b
\right)
\eeq
where the axial current ${\cal A}^\mu$ is
\beq
{\cal A}^\mu =\frac{i}{2}\left(\xi^\dagger
\partial^\mu\xi-\xi\partial^\mu\xi^\dagger\right)
\eeq
and $\xi={\rm exp}(i{\cal M}/f),$ with ${\cal M}$
being the usual $3\times 3$ matrix describing the
octet of pseudoscalar Nambu-Goldstone bosons. The
axial coupling constant $g$ is one of the basic
parameters of HM$\chi$L, which is of direct import
to our problem. $a,b$ denote light quark flavours
($a,b$=1,2,3) and f is the pion decay constant,
$f=132$ MeV. Expanding the axial current and using
the first term ${\cal A}^\mu=-(1/f)\partial^\mu
{\cal M}+....$ one obtains the effective
Lagrangian representing $B^*B$-pion and
$B^*B^*$-pion interactions, which are the relevant
ones in our problem. Thus,
\beq
\label{leff}
{\cal L}^1_{\rm eff}=
\left[-\frac{2g}{f}B^*_\mu \partial^{\mu}{\cal M} B^{\dagger}
+h.c.\right]+\frac{2 g
i}{f}\epsilon_{\alpha\beta\mu\nu}{B^*}^{\beta}
\partial^{\mu}{\cal M} {{B^*}^{\dagger}}^\alpha v^\nu .
\eeq

The dimensionless ${B^*}B\pi$ coupling is defined
as \cite{Casalbuoni}
\beq
\label{defgb}
\langle \pi(q)\bar{B}(v_1)|{B^*}(v_2,\epsilon_2)
\rangle = g_{B^{*} B\pi}(q^2) q_\mu \epsilon_{2}^{\mu}
\eeq
where $\epsilon^\mu$ is the polarization vector of
$B^{*},$ with the physical coupling given by the
limit $q^2\rightarrow m_\pi^2.$ We use the same
normalization convention as in  \cite{Casalbuoni}.

Throughout this work, we assume that the variation
of $g_{B^{*} B\pi}$ with $q^2$ in the region of
our treatment can be safely neglected. Likewise,
we define
\beq
\langle \pi(q)\bar{B^*}(v_1,\epsilon_1)|B^*(v_2,\epsilon_2)
\rangle = g_{B^{*} B^* \pi}(q^2)
\epsilon_{\alpha\beta\mu\nu}\epsilon_1^\alpha \epsilon_2^\beta
q^\mu v_1^\nu,
\eeq
with the same remarks as above.

We also note that the isospin symmetry requires
\beqa
\label{gcopb}
g_{B^* B\pi} & \equiv & g_{{B^*}^+ B^0\pi^+} =
-\sqrt{2}g_{{B^*}^+ B^+\pi^0} = \sqrt{2}g_{{B^*}^0 B^0\pi^0}
\nonumber \\
&=& -g_{{B^*}^0 B^+\pi^-}
\eeqa
and the $g_{B^* B\pi} $ so defined is the commonly
used in the literature.

The (\ref{gcopb}) relation holds similarly for $g_{B^*
B^*\pi},g_{D^* D\pi}$ and $g_{D^* D^*\pi}$
couplings.

Now, from (\ref{lchir})-(\ref{leff}), using the
definition  (\ref{defgb}) one has
\beqa
\label{gchir}
g_{B^* B\pi} &=&\frac{2 M_B}{f}g=g_{B^* B^*\pi}
\nonumber \\
g_{D^* D\pi} &=&\frac{2 M_D}{f}g=g_{D^* D^*\pi}.
\eeqa
Note that, in deriving (\ref{gchir}) one assumes
$B,B^*(D,D^*)$ mass degeneracy.  In order to
calculate the \Bsgg  decay width we use the
interaction Lagrangian (\ref{leff}) to the leading
order in chiral perturbation theory, which is an
appropriate tool here, in view of the smallness of
$M_{B^*}-M_B.$

The calculation of the radiative processes
$B^*(D^*)\rightarrow B(D)\gamma\gamma$ obviously
requires the incorporation of the electromagnetic
interaction in our  Lagrangian (\ref{leff}), which
is performed \cite{Stewart,Cheng2} by the usual
procedure of gauging the Lagrangian with the
$U(1)$ photon field. This leads to the replacement
of the derivative operators in the Lagrangian by
covariant derivatives containing the photon field,
explicitly  exhibited in \cite{Cheng2}.
Nevertheless, the new Lagrangian still does not
provide for couplings to induce the observed \Bsg,
\Dsg magnetic dipole transitions. This necessitates
the introduction of an additional term in the
Lagrangian, a contact gauge invariant interaction
proportional to the electromagnetic field
$F_{\mu\nu},$ which is given by
\cite{Stewart,Cheng2}

\beq
\label{lem}
{\cal L}= \frac{e\mu}{4} {\rm Tr}
\left( \bar{H}_{a} \sigma_{\mu\nu} F_{\mu\nu}H_{b}\delta_{ab}
\right),
\eeq
where $\mu$ is the strength of this anomalous
magnetic dipole interaction, having mass dimension
$[1/M].$

Additional terms arising from an $1/M_{H}$
expansion exist \cite{Stewart,Cheng2}, however
several of them, including the radiation of the
heavy quark can be absorbed in equation
(\ref{lem}) by redefining $\mu.$ In the present
paper, we shall consider $\mu$ as an effective
coupling, representing the strength of the
$B^*(D^*)B(D)\gamma$ transition.

Expanding $H$ in terms of the components, one
obtains for the additional electromagnetic
interaction

\beq
\label{leffem}
{\cal L}^2_{\rm eff}=-e\mu F^{\mu\nu}
\left[i{B^*_\mu}^{\dagger}B^*_\nu
+\epsilon_{\alpha\beta\mu\nu}v^\alpha
(B^{\dagger}B^*_\beta+h.c.)\right]
\eeq

which exhibits  $B^*B\gamma$ and  $B^*B^*\gamma$
couplings with equal strength, as given by the
heavy quark symmetry. From (\ref{leffem}) we
obtain the respective vertices, which are
\beqa
\langle \gamma(k,\epsilon) \bar{B^*}(v_{1},\epsilon_{1})
|B^*(v_{2},\epsilon_{2})\rangle&=& e \mu
M_{B^*}(\epsilon_{1}.k\epsilon.\epsilon_{2}-
\epsilon_{2}.k\epsilon.\epsilon_{1})
\\
\langle \gamma(k,\epsilon) \bar{B}(v_{1})
|B^*(v_{2},\epsilon_{2})\rangle &=&
-i e M_{B^*}\mu \epsilon_{\mu\nu\alpha\beta}{\epsilon}^{\mu}
k^{\nu}v_{2}^{\alpha}\epsilon_2^{\beta}.
\eeqa

The propagator of the heavy vector meson is given
by $-i(g^{\mu\nu}-v^\mu
v^\nu)/2\left[(v.k)-\Delta/4\right],$ where
$\Delta = M_{B^*}-M_B$ and $v,k$ are the velocity
and the residual momentum. The propagator of the
heavy pseudoscalar meson is
$i/2\left[(v.k)+3\Delta/4\right]$ \cite{Casalbuoni}.

Now, considering Lagrangians (\ref{leff}) and
(\ref{leffem}), as well as the axial anomaly
responsible for the $\pi^0\rightarrow
\gamma\gamma,$ we classify the diagrams contributing
to \Bsgg in the leading order of chiral
perturbation theory as follows:

There is the diagram ${B^*}^0\rightarrow B^0
``\pi"\rightarrow B^0
\gamma \gamma$ (Fig.1), via a virtual pion (the
somewhat different situation in \Dsgg will be
analyzed in the last section). This diagram
contains  the known strength of the pion axial
anomaly. Then, there is the loop graph
${B^*}^0\rightarrow ({B^*}^{+}
\pi^{-})\rightarrow B^0 \gamma \gamma$  (Fig.2), with the
photons radiated from the virtual charged pion in
the loop, with additional graphs of the same
class, as specified in the next section. The first
graph is proportional to the $g_{B^* B\pi}$
coupling, while the loop graph contains the
$g_{B^* B\pi} g_{B^* B^*\pi}$ product. In addition
we have the tree level diagram with two insertions
of the magnetic operator defined in
(\ref{leffem}),leading to ${B^*}^0\rightarrow
{B^*}^0\gamma\rightarrow B^0\gamma \gamma$
(Fig.3). This graph depends only on the $\mu$
magnetic moment. Finally we have a class of one
loop diagrams which involve both the magnetic
moment and the strong coupling, which is exhibited
in Fig.(4-6). We did not include the contribution of
diagrams containing three heavy meson propagators
which is negligible.
 Needless to say, the determination of $g_{B^*
B\pi}$ would be simpler, should the first two
graphs dominate. However, this  is not true for
$D^*$ decay, while it can be true for the $B^*$
decay for an opportune range of parameters as we
discuss in the next section.

We remark at this point that corrections to
(\ref{lchir}) which arise from higher terms in the
$1/M_H$ expansion as well as in chiral breaking
 have also been investigated
\cite{Stewart,Casalbuoni,Cheng,Boyd}. A
comprehensive inclusion of these corrections in
the calculation of the two-photon decays of heavy
vector mesons is beyond our scope in this first
treatment of these processes. Nevertheless, we
note that we shall use physical masses for the
degenerate doublet of heavy mesons both in the
loop propagators and in the decay calculations,
moreover the chiral loops included are themselves
of order $1/M_{H}.$ The terms we include are the
leading ones in chiral perturbation theory, and
are of the same order in an $1/N_c$ expansion;
moreover to this order there are no counterterms
\cite{Leibovich}.

It is appropriate to mention now that the $g_{B^*
B\pi},g_{D^* D\pi}$ couplings were estimated in
recent years by the use of a variety of
theoretical techniques, like QCD sum rules
\cite{Dosch,Belyaev}, soft pion approximation
\cite{Cola} and other methods
\cite{Cho,Amundson,Stewart,Cheng,Boyd,Manohar}.
Generally, the values of $g$ obtained in these
works are in the range $g=0.25-0.7,$ significantly
smaller than the quark model result of $g=1$ or of
modified quark models \cite{Pham,Isgur,Nussinov}
which brought this value slightly below one. The
most recent determinations of $g$ include  an
analysis \cite{Casalbuoni} of various theoretical
approaches which leads to a ``best estimate" of
$g=0.38,$ a recent lattice determination giving
$0.42(4)(8)$ \cite{UKQCD} and the analysis of
Stewart \cite{Stewart} which incorporates symmetry
breaking terms in the Lagrangian and obtains
$g=0.27^{+0.9}_{-0.4}.$

Finally, the experimental limit
$\Gamma({D^*}^+)<131$ KeV \cite{ACCMOR} puts an
upper limit of $g<0.71,$ using
$\Gamma({D^*}^+\rightarrow D^0\pi^+ +D^+\pi^0)=
(g^2/4\pi f^2) |\vec{p}_\pi|^3.$

The existing theoretical estimates we mentioned,
give $0.04 {\rm KeV} <\Gamma(\Bosg)< 1 {\rm KeV}$
and $0.10 {\rm KeV} <\Gamma(\Bpsg)< 1 {\rm KeV},$
where we allowed for a slightly higher upper limit.
We redefine the magnetic coupling in
eq.(\ref{leffem}) to a dimensionless quantity
$\bar{\mu}=M_{B^*}\mu=g_{{B^*}^0B^0\gamma},$ and
$\bar{\mu}_{+}=M_{B^*}\mu_{+}=g_{{B^*}^+B^+\gamma},$
then the above limits give the following ranges,
$2.2<\bar{\mu}<11.0,$ and $3.5<\bar{\mu}_+<11.0.$

\section{The decay amplitudes.}

We present now the explicit expressions of the
decay amplitudes, which in our approach, to
leading order in chiral perturbation theory,
consist of the contribution of the anomaly graph
(Fig. 1), the tree level  graph (Fig.3) and the
loop graphs (Figs. 2,4-6).

In presenting the differential decay distribution,
we use the following variables:
\beqa
s&=&(p-p^\prime)^2=(k_1+k_2)^2\nonumber \\
t&=&(p-k_1)^2=(p^\prime+k_2)^2\nonumber \\
u&=&(p-k_2)^2=(p^\prime+k_1)^2
\eeqa
with
\beq
t+s+u=M_{B^*}^2+M_{B}^2,
\eeq
where  $k_1, k_2$ are the four momenta of the two
photons and $p,p^\prime$ are the four-momenta of
the decaying $B^*$ and the final $B$ respectively.
The allowed ranges for $s$ and $t$ are
\beqa
0\leq s&\leq& (M_{B^*}-M_{B})^2,
\,\,\,\,\,t_{-}\leq t\leq t_{+}\nonumber \\ t_{\pm}
&=&
\frac{1}{2}\left[(M_{B^*}^2+M_{B}^2-s)\pm
\sqrt{(M_{B^*}^2+M_{B}^2-s)^2-4M_{B^*}^2M_{B}^2}\right].
\eeqa

 The amplitudes are given for
\Bosgg and we shall remark on the changes
appearing in \Dosgg whenever required. The
amplitude from the anomaly graph, mediated by a
pion is:

\beqa
\label{Ampan}
{\rm A}_{\rm anomaly}(\Bosgg)=\frac{\alpha
g_{B^*B\pi}}{\sqrt{2}\pi f} \epsilon^\mu_{B^*}
\epsilon_{1}^\lambda \epsilon_{2}^\gamma
\frac{1}{s -m_\pi^2}
\epsilon_{\lambda\gamma\tau\rho}k_1^\tau k_2^\rho (k_1+k_2)_\mu,
\eeqa
where  $\epsilon_{B^*},
\epsilon_1, \epsilon_2$ are the polarization
vectors of the heavy vector meson ${B^*}$ and the
two photons respectively. There are additional
contributions from $\eta,\eta^{\prime}$ which are
not specified in (\ref{Ampan}). As we shall
describe in the next section, their contribution
is rather small and we may safely neglect them at
this stage.

For the tree level graph we find
\beqa
\label{Atree}
{\rm A}_{\rm tree}(\Bosgg)&=&
\frac{4\pi \alpha \bar{\mu}^2}{ M^{*}_{B}}
\epsilon_{\gamma\delta\alpha\beta}{\epsilon_{B^*}}_ {\sigma}
{p^{\prime}}^\alpha \nonumber \\ && \left[\frac{
\epsilon_{2}^\gamma k_2^{\delta}(\epsilon_{1}^{\sigma}
k_{1}^{\beta}-\epsilon_{1}^{\beta}
k_{1}^{\sigma})}{t-M_{B^*}^2}+
\frac{
\epsilon_{1}^\gamma k_1^{\delta}(\epsilon_{2}^{\sigma}
k_{2}^{\beta}-\epsilon_{2}^{\beta}
k_{2}^{\sigma})}{u-M_{B^*}^2}\right]
\eeqa

The loop contribution  which depends only on the
strong coupling is given by a sum of 
several
diagrams. In addition to that explicitly shown in
Fig.2 there are diagrams with one photon radiated
by the virtual pion in the loop  and the other
emitted from the $B^* B^*
\pi,$ $B^* B \pi$ vertices or both photons emitted
from these vertices, or both photons emitted from
the loop by the $\pi\pi\gamma\gamma$ vertex.
In the limit of photon momenta small compared to
the pion mass, which we find to be a suitable
approximation, the class of diagrams of Fig.2
gives
\beqa
\label{Amploop1}
{\rm A}^{1}_{\rm loop}(\Bosgg) &=& \frac{\alpha
g_{B^* B\pi} g_{B^* B^*\pi}}{8\pi M_{B^*}}
\epsilon_{\mu\eta\alpha\beta} \epsilon^\mu_{B^*} v^{\beta}
\epsilon_{1}^\lambda \epsilon_{2}^\gamma (k_1+k_2)^\eta
\nonumber \\
& & \times
\left[3\,\left( g^\alpha_\gamma v_\lambda +
g^\alpha_\lambda v_\gamma \right)+\frac{1}{9
m_{\pi}}\left( g^\alpha_\gamma {k_2}_\lambda +
g^\alpha_\lambda {k_1}_\gamma \right)\right].
\eeqa

Finally other loop contributions to the \Bsgg
decay come from diagrams where both the strong
coupling and the magnetic one are involved. There
are diagrams with one photon radiated by the
virtual pion in the loop and the other emitted
from the ingoing $B^*$ particle through the
$B^*B^*(B)\gamma$ vertices (Fig.4), or from the
outgoing particle $B^*$ which becomes $B$ through
the $B^*B\gamma$ vertex (Fig.5), or from the $B$
in the loop which becomes $B^*$ through the
$B^*B\gamma$ vertex (Fig.6).

The amplitude corresponding to Fig.4 with a
$B^*B^*\gamma$ vertex is:
\beqa
\label{Amploop2}
{\rm A}^{2}_{\rm loop}(\Bosgg) &=&\frac{11\alpha
(M_{B^*}-M_B)g_{B^* B^*\pi}g_{B^*
B\pi}\bar{\mu}}{16\pi
M^2_{B^*}}\epsilon_{\alpha\sigma\gamma\delta}\nonumber
\\
&& \left[(k_1^\alpha(\epsilon_{B^*}.\epsilon_1)-
\epsilon_1^\alpha(\epsilon_{B^*}.k_1))
\frac{(p-k_1)^\delta\epsilon_2^\gamma k_2^\sigma}{t-M^2_{B^*}}
\right.
\nonumber \\
&+& \left.
(k_2^\alpha(\epsilon_{B^*}.\epsilon_2)-
\epsilon_2^\alpha(\epsilon_{B^*}.k_2))
\frac{(p-k_2)^\delta\epsilon_1^\gamma k_1^\sigma}{u-M^2_{B^*}}
\right]
\eeqa

The amplitude corresponding to Fig.4 where a
$B^*B\gamma$ vertex replaces the $B^*B^*\gamma$
appearing in Fig.4, is
\beqa
\label{Amploop3}
{\rm A}^{3}_{\rm loop}(\Bosgg) &=&\frac{9\alpha
m_{\pi}^2 g_{B^* B\pi}^2\bar{\mu}}{4\pi
M_{B^*}}\epsilon_{\mu\nu\alpha\beta}\epsilon^{\nu}_{B^*}
v^{\alpha}v_{\gamma}\nonumber
\\
&&
\left[\frac{\epsilon_1^{\mu}k1^{\beta}\epsilon_2^{\gamma}}
{t-M^2_{B^*}}+
\frac{\epsilon_2^{\mu}k2^{\beta}\epsilon_1^{\gamma}}
{u-M^2_{B^*}}
\right]
\eeqa

The amplitude corresponding to Fig.5 with $B$ in
the loop is:
\beqa
\label{Amploop4}
{\rm A}^{4}_{\rm loop}(\Bosgg) &=&
\frac{3 \alpha m_{\pi}^2 g_{B^* B\pi}^2\bar{\mu}}{8 \pi M_{B^*}^2}
\epsilon_{\mu\nu\alpha\rho} \epsilon^\sigma_{B^*}
(g_{\gamma}^{\rho}v_{\sigma}+g_{\sigma}^{\rho}v_\gamma)
\nonumber \\
&&
\left[
\frac{(p-k_1)^\alpha\epsilon_2^\mu k_2^\nu\epsilon_1^{\gamma}}
{t-M^2_{B^*}}+
\frac{(p-k_2)^\alpha\epsilon_1^\mu k_1^\nu\epsilon_2^{\gamma}}
{u-M^2_{B^*}}
\right]
 \eeqa

 and the one with $B^*$ in the loop is
 \beqa
\label{Amploop5}
{\rm A}^{5}_{\rm loop}(\Bosgg) &=&
\frac{3 \alpha m_{\pi}^2 g_{B^* B^*\pi}^2\bar{\mu}}{16 \pi M_{B^*}^3}
\epsilon_{\alpha\beta\gamma\delta} \epsilon^\beta_{B^*}v^{\delta}
\epsilon_{\eta\alpha\xi\rho}\epsilon_{\mu\nu\sigma\eta}
(g^{\gamma\rho}v_{\tau}+g^{\gamma}_{\tau}v^\rho)
\nonumber \\
&&
\left[
\frac{(p-k_1)^\xi(p-k_1)^\sigma\epsilon_1^\tau
\epsilon_2^\mu k_2^\nu}{t-M^2_{B^*}}+
\frac{(p-k_2)^\xi(p-k_2)^\sigma\epsilon_1^\mu
\epsilon_2^\tau k_1^\nu}{u-M^2_{B^*}}
\right]
 \eeqa

 Finally the amplitude corresponding to Fig.6 is:
 \beqa
\label{Amploop6}
{\rm A}^{6}_{\rm loop}(\Bosgg) &=&
\frac{11 \alpha  g_{B^* B\pi}^2 \bar{\mu}_{+}}{24 \pi M_{B^*}^3}
\epsilon_{\mu\nu\alpha\beta}
\epsilon^\sigma_{B^*}
\left[\frac{1}{3}(k_1+k_2)_\sigma
\epsilon_1^\alpha\epsilon_2^\mu k_1^\nu k_2^\beta
\right.
\nonumber \\
&+&
\left.
\frac{1}{2}p^\alpha\left[\epsilon_2^\mu k_2^\beta
({\epsilon_1}_\sigma k_1^\nu+\epsilon_1^\nu
{k_1}_\sigma)+
\epsilon_1^\mu k_1^\beta
({\epsilon_2}_\sigma k_2^\nu+\epsilon_2^\nu
{k_2}_\sigma)\right]\right].
\eeqa

In the $B^*$ decay, the pions are the sole
contributions in the loop, while in the \Dsgg
calculation we include both pions and kaons.

Let us call A the sum of all the amplitudes:
\beq
{\rm A}={\rm A}_{\rm anomaly}+{\rm A}_{\rm tree}+
{\rm A}_{\rm loops}
\eeq
where ${\rm A}_{\rm loops}$ is the sum of all the
amplitudes  $A^{i}_{\rm loops},\,i=1,...,6$ which
come from the loops.

The square of the above amplitude, when averaged
over the initial spin and summed over the final
spins is:
\beq
|\bar{A}|^2=\frac{1}{2}\sum_{\rm spins} |A|^2,
\eeq

where we have included the factor $\frac{1}{2}$ in
order to take into account two identical particles
in the final state.

The differential decay rate of photon energy is
obtained by integrating the following expression
over the variable $t,$

\beq
\label{dgamma}
\frac{d\Gamma}{ds}=\frac{1}{(2\pi)^3}\frac{1}{32 M_{B*}^3}
\int_{t_{-}}^{t_{+}} |\bar{A}|^2 dt.
\eeq

There is a major difference in the anomaly
contribution of the $B^*$ and $D^*$ decays. Since
the $\pi^0$ appears in the physical region in the
\Dosp decay we have to isolate the on-shell $\pi^0$ decay
in the \Dosgg mode. Hence, for the $D^*$ case we
limit ourselves in the integration of $ d\Gamma/ds
$ to a region which goes from $s
=0$ up to $20$ MeV away from the pion mass.

Using now the physical masses of
 $ M_{B^*}, M_{B}, M_{D^*}
M_{D}$ \cite{PDG} and the eq.(\ref{gchir}) we
obtain for the decay rates of
\Bsgg, the following expression:
\beqa
\label{gbsgg}
\Gamma(\Bsgg) &=& 3.40\times 10^{-16}  g^2 + 1.53\times 10^{-12} g^4 +
   4.81\times 10^{-17} g^3 \bar{\mu} \nonumber \\
   &+& 1.53\times 10^{-13} g^4 \bar{\mu}+
   4.71\times 10^{-17} g \bar{\mu}^2 + 9.81\times 10^{-14} g^4 \bar{\mu}^2
   \nonumber \\
   &+&
   2.65\times 10^{-16} g^2 \bar{\mu}^3 + 1.38\times 10^{-16} \bar{\mu}^4+
   2.67\times 10^{-16} g^3 \bar{\mu}_{+}
   \nonumber \\
   &+&    2.90\times 10^{-17}g^4 \bar{\mu}_{+} +
   8.94\times 10^{-20} g^4 \bar{\mu} \bar{\mu}_{+}
   \nonumber \\ &+&
   9.11\times 10^{-20} g^2 \bar{\mu}^2 \bar{\mu}_{+}+
   7.21\times 10^{-17} g^4 \bar{\mu}_{+}^2.
\eeqa
As we can see this is a function of both the
strong coupling $g$ and the magnetic dipole
strength $\bar{\mu},\bar{\mu}_{+}$ which represent
the effective $g_{{B^{*}}^0B^0\gamma}$ and
$g_{{B^{*}}^+B^+\gamma}$ couplings. For the
\Dsgg we have:
\beqa
\label{gdsgg}
\Gamma(\Dsgg) &=& 2.52\times 10^{-11} g^2 + 5.85\times 10^{-10} g^4 +
   1.79\times 10^{-12} g^3 \bar{\mu}
   \nonumber \\
   &+& 4.43\times 10^{-11} g^4 \bar{\mu} +
    1.30\times 10^{-12} g \bar{\mu}^2 + 3.50\times 10^{-11}
g^4 \bar{\mu}^2 \nonumber \\ &+&
   2.42\times 10^{-13} g^2 \bar{\mu}^3 + 2.18\times 10^{-12} \bar{\mu}^4 +
   1.01\times 10^{-11} g^3 \bar{\mu}_{+}
   \nonumber \\
   &+& 2.95\times 10^{-13} g^4 \bar{\mu}_{+} +
   2.11\times 10^{-14} g^4 \bar{\mu} \bar{\mu}_{+}
   \nonumber \\
   &+&
   1.70\times 10^{-14} g^2 \bar{\mu}^2 \bar{\mu}_{+} +
   2.05\times 10^{-12} g^4 \bar{\mu}_{+}^2.
   \eeqa
In this  case we can relate the magnetic coupling
to the strong coupling
 using the existing
experimental informations on $\Gamma(\Dosp):
\Gamma(\Dosg)$ of $(61.9\pm 2.9)\%:(38.1\pm 2.9)\%,$
which gives rise to the relation: $ \bar{\mu} \simeq 6.6
g,$ and $\Gamma(\Dpsp):
\Gamma(\Dpsg)$ of $(67.6\pm 0.9)\%:(1.7\pm 0.6)\%,$
which gives rise to the relation: $ \bar{\mu}_{+}
\simeq 1.7 g.$ Then we can write the decay width
solely as a function of  $g,$ which is a crucial
step in the engagement of this decay as a tool for
measuring $g.$

\beqa
\label{pos}
\Gamma(\Dsgg)&=&
2.52\times 10^{-11} g^2 + 5.66\times 10^{-11} g^3
\nonumber \\
&+& 4.76\times 10^{-9} g^4 +
   3.64\times 10^{-10} g^5 + 1.53\times 10^{-9} g^6.
   \eeqa

We used in eqs.(\ref{gbsgg})-(\ref{gdsgg}) the
same notation of $\mu,\mu_+$ for the magnetic
moment strength in both the charmed and beauty
sectors although they are probably not equal for
the physical processes. However since in the charm
case, the magnetic coupling  has been related to
the strong one, the $\mu, \mu_+$ will denote in
the rest of the paper the strength of the
${B^*}^{0,+}\rightarrow B^{0,+}\gamma$
transitions.

The experimentally measured branching ratios of
\Dsp, \Dsg lead to relations between $\bar{\mu},
\bar{\mu}_+$ and $g$ modulo an unknown phase.
We have allowed also for the possibilities of
negative relative signs among $\bar{\mu},
\bar{\mu}_+$ and $g$ and as it turns out, this
affects only slightly the numerical picture, due
to the fact that the main contribution is given by
quadratic terms. In eq.(\ref{neg}) we give for
comparison the expression obtained for
$\bar{\mu}=-6.6 g,
\bar{\mu}_+=-1.7 g.$
\beqa
\label{neg}
\Gamma(\Dsgg)_{\bar{\mu}={\rm neg}}&=&
2.52\times 10^{-11} g^2 + 5.66\times 10^{-11} g^3
\nonumber \\
&+& 4.70\times 10^{-9} g^4 -
   3.64\times 10^{-10} g^5 + 1.53\times 10^{-9} g^6.
   \eeqa

This ambiguity will be further discussed in the
next section.
 The difference between the rates of $B^*$ and
$D^*$ is mainly due to the different phase space.

In discussing the two photon radiative decays, we
shall refer in the next section to the following
quantities

\beq
\label{BRb}
{\rm BR}(\Bsgg) =
\frac{\Gamma(\Bosgg)}{\Gamma({B^*}^0)}=
\frac{\Gamma(\Bosgg)}{\Gamma(\Bosg)}
\eeq
\beq
\label{BRd}
{\rm BR}(\Dsgg) =
\frac{\Gamma(\Dosgg)}{\Gamma({D^*}^0)}=
\frac{\Gamma(\Dosgg)}{\Gamma(\Dosg)+\Gamma(\Dosp)}.
\eeq

\section{Discussion and Conclusions.}

The formalism we have presented refers to the
decays of the neutral heavy vector mesons
${B^*}^0,{D^*}^0,$ as it will the numerical
analysis of our results, which will be given
below. For the charged decays, \Bpsgg and \Dpsgg
one has to consider also the bremsstrahlung
emission which appears in diagrams of Fig.3 and
Fig.4 and additional ones.The bremsstrahlung
radiation comes from the initial or the final
charged particles. To give an idea of this effect
we have calculated the part of the bremsstrahlung
amplitude which is due to radiation from the final
$B^+$ particle in the amplitude
${B^*}^+\rightarrow (B^+\gamma)\rightarrow
B^+\gamma\gamma.$ It is
\beq
\label{Ampbrem}
A^{B^+}_{\rm  brem}= 4\pi\alpha g_{B^* B\gamma}
\epsilon^\mu_{B^*}
\epsilon_{1}^\lambda \epsilon_{2}^\gamma p^\alpha
\left[
\frac{ \epsilon_{\lambda \mu \alpha\beta} k_1^\beta
p^\prime_\gamma}{(t-M_B^2)} +
\frac{\epsilon_{\gamma \mu \alpha\rho} k_2^\rho
p^\prime_\lambda}{(u-M_B^2)}
\right].
\eeq

This amplitude (and the ones given by ${B^*}^+$
radiation) have to be added  in order to get the
full amplitude for the \Bpsgg decay. An estimate
of the bremsstrahlung decay width, from
(\ref{Ampbrem}) only, using for the unknown
$g_{B^*B\gamma}$ vertex a value leading to
$\Gamma(\Bpsg)=0.14$ KeV \cite{Casalbuoni} leads
to a decay width of $\sim 10^{-9}$ GeV for
$k_1,k_2\geq 10$ MeV, considerably larger than
 (\ref{gbsgg}). The use of the charged
\Bpsgg thus involves a different type of analysis
in view of the relative size of the different
components of $|A(\Bpsgg)|^2$ and is less useful
for a determination of $g.$ A similar situation is
encountered for
\Dpsgg. Thus, we concentrate here on the ``safer''
neutral decays and we relegate the discussion of
the charged decays to a separate work, in which we
consider  the usefulness of $\Gamma(\Bpsgg)$ for
the determination of $g_{{B^*}^+B^+\gamma} $
\cite{Dafne}.

We proceed now to analyse the results on the two
decays separately and we start with \Dosgg
transition for which the rate (\ref{pos}) was
obtained.

The many theoretical estimates for $g$ we
mentioned in Sec.3 are spread over the range
$0.25<g<1$ (we also remind the reader that the
experimental result \cite{ACCMOR} on the upper
limit of $\Gamma({D^*}^+\rightarrow all)$ can be
interpreted as $g<0.71$). Using this range and,
eqs.(\ref{pos}),(\ref{neg}) we can establish the
expectation
\beq
\Gamma(\Dosgg)\simeq(0.022-6.73){\rm eV}.
\eeq
The most promising feature of the present analysis
arises when we use the existing experimental
informations on $\Gamma(\Dosp):\Gamma(\Dosg)$ of
$(61.9\pm 2.9)\%:(38.1\pm2.9)\%$ to transform
eq.(\ref{BRd}) into a ratio of $\Gamma(\Dosgg)$ to
the total ${D^*}^0$ width which becomes
proportional to $g^2.$

 Using
\beq
\Gamma(\Dosp)=\frac{1}{12\pi}\frac{g^2}{f^2}
|\vec{p}_\pi|^3=1.25\times 10^{-4} g^2 {\rm GeV}
\eeq
and the $(61.9\pm 2.9)\%:(38.1\pm 2.9)\%$
relative branching ratio, we arrive at
\beq
\label{gamma}
\Gamma({D^*}^0\rightarrow all)=
(2.02\pm 0.12)\times 10^{-4} g^2 {\rm GeV}.
\eeq
Thus from (\ref{gamma}) with (\ref{pos}) we can
obtain a  branching ratio {\it which depends on}
$g^2$ {\it only:}

\beq
\label{gd}
{\rm Br}(\Dosgg)= \frac{(0.025+ 0.057 g + 4.76 g^2
+
   0.36 g^3 +1.53 g^4)\times 10^{-9}g^2}
{2.02\times 10^{-4} g^2}.
\eeq
With our model for \Dosgg,  the measurement of
this ratio  will thus constitute a measurement of
the $g$ coupling. Using again the accepted
expectation of $0.25\leq g\leq 1,$ we predict
\beq
\label{bdsgg}
{\rm
Br}(\Dsgg)=\frac{\Gamma(\Dosgg)}{\Gamma(D^*)}=
(0.16-3.3)\times 10^{-5}.
\eeq

A few remarks are in order. Firstly, the sign
question The observed branching ratios do not
afford to estabilish experimentally the sign of
$g/g_{{D^*}^0D^0\gamma}.$ On the other hand, there
is theoretical support from the analysis of
Stewart \cite{Stewart} on the positive sign of
this ratio. However even if we assume opposite
sign for various pairs of the couplings, we found
that  the changes are rather small, and this is
explicitly exhibited in Table 2, and included in
Eq.(\ref{bdsgg}).

The differential distribution in the $s$-variable
can also be used to learn about the value of $g,$
due to the fact that the different contributions
depend on different powers of $g.$ Finally we
remark that the contributions from the diagrams
exhibited in Fig.4 are rather small, as a result
of two heavy propagators. The main contributions
are those of the anomaly and of the graphs of
Fig.(2) and (3). In Figs.(7) and (8) we present
the differential distributions in $s$ for $g=0.7$
and $g=0.25.$ In the latter, the contributions
containing a higher power of $g$ are diminished
and the effect of the anomaly becomes visible in
the higher end of the spectrum.

Turning now to the \Bsgg, we have a rather
different situation. Firstly, there is only one
major decay of $B^*,$ namely \Bsg, which precludes
an analysis like in $D^*$ decays. The branching
ratio (\ref{BRb}) depends on three parameters,
$g_{B^*B\pi}$ (or $g$), $g_{{B^*}^0B^0\gamma}$ (or
$\mu$) and $g_{{B^*}^+B^+\gamma}$ (or $\mu^+$). At
this point, we rely on the theoretical estimates
presented in Section 2 and contain our analysis to
the regions given by existing calculations.

Now, an inspection of eq.(\ref{gdsgg}) shows that
$\mu_+,$ which appears only in diagram of Fig.6
has a very little effect on the rate, whether
$\mu,\mu_+$ are at the lowest or at the highest
end of their value, for any value of $g.$ Hence,
we continue our analysis in the parameter space of
$[g,\mu]$ only.

In table 2 we present the values of ${\rm
Br}(\Bsgg)$ for different values of $g$ and the
two extreme values of $\mu,$ corresponding to
$\Gamma(\Bsg)=40 {\rm eV}$ and $1{\rm KeV}.$
Again, assuming that relative negative signs are
possible we give in the last column the Br for
$\mu=-2.2.$ Clearly, a branching ratio in the
$10^{-7}-10^{-6}$ range will not allow to pinpoint
accurate values for the two couplings.

Nevertheless if the branching ratio turns out to
be in the $10^{-5}$ range, it can only be caused
by large values of $g,$ say $g>0.6.$

We wish also to mention an additional scenario:
the $g$ coupling will probably be measured
directly in $D^*$ decays, or indirectly from ${\rm
Br}(\Dosgg)$ or other methods. With this
knowledge, ${\rm Br}(\Bosgg)$ becomes a function
of $g_{{B^*}^0B^0\gamma}$ only and it could
provide the desirable measurement of this
coupling. This is a very interesting issue, since
as pointed out already some time ago
\cite{Miller}, there is no other possibility of
measuring the width of the \Bsg decay with
presently known techniques.

In Figs.(9),(10) and (11) we give the differential
distribution of $d\Gamma(\Bsgg)/ds$ for $g=0.5$
and three different values of $\mu.$ Clearly, once
$g$ is known one may use accurate differential
distributions to distinguish between different
$\mu$ values.

At this point we also wish to make some remarks on
the similar decays of the strange heavy vector mesons,
${B_s^{*}}^0 \to \, B^0_s\, \gamma\,\gamma$ and
${D_s{*}}^+ \to \, D^+_s\, \gamma\,\gamma$ which
were not mentioned so far. In both these cases the
pion anomaly is further suppressed, since
${B_s^{*}}^0 \to \, B^0_s\, \pi^0,
{D_s^{*}}^+ \to \, D^+_s\, \pi^0  $ can proceed
only by isospin violation, e.g. via $\eta-\pi^0$
mixing.
On the other hand, both decays can proceed via chiral
loops with charged $K$-mesons in the loops,
${B^{*}}^0_s\to(K^+ {B^*}^-) \to \, B^0_s\, \gamma\,\gamma$ and
${D^{*}}^+_s \to(K^+ {D^*}^0)\to \, D^+_s\, \gamma\,\gamma.$
However, one must add that for
${D^{*}}^+_s \to \, D^+_s\, \gamma\,\gamma$
there is the complication of the bremstrahlung
and we shall disregard it here.

We calculated therefore only  the ${B^*}^0_s \to B^0_s \gamma
\gamma$ decay, and the situation is quite similar to that encountered
in ${B^*}^0$ decay; therefore we do not repeat
this analysis here.

Before concluding we comment on
a few points which were neglected in our treatment.
\begin{itemize}
\item[1]
We calculated also the contribution to the anomaly
term of a virtual $\eta$ exchange for the \Dsgg
decay. The inclusion of $\eta$ modifies our result
in eq.(\ref{gdsgg}) by a factor of
$\left(1+\frac{g_{D^* D\eta}}{10
g_{D^*D\pi}}\right).$ Since $g_{D^* D\eta}$ and $
g_{D^*D\pi}$ are expected to be comparable, this
is a small effect.
\item[2]
We neglected the off-shell $q^2$-dependence of the
anomaly which could have some effect, especially in
the $D^*$-decay. This should be included in a
more detailed treatment.
\end{itemize}

To summarize, we have used the Heavy Meson Chiral
Lagrangian to present the first treatment of the
rare \Bosgg, \Dosgg decays. The decay rates depend
on the strong $g_{\cbsbp},g_{\cdsdp}$ couplings
and on the strength of the magnetic dipole
transitions  $g_{B^*B\gamma},g_{D^*D\gamma}.$ The
strong couplings are expressed in the chiral
Lagrangian by the strong axial coupling $g.$

We have shown that Br(\Dosgg) can be given as a
function of $g$ only, and as such, it would
provide an appropriate tool for its measurement.
For the conventionally envisaged range $0.25<g<1$
we calculated $1.6\times 10^{-6}<{\rm
Br}(\Dosgg)<3.3\times 10^{-5}.$

On the other hand, Br(\Bosgg) is a function of
$g_{\cbsbp},g_{{B^*}^0B^0\gamma}$ and
$g_{{B^*}^+B^+\gamma}.$ The latter coupling has
little effect on the branching ratio.
Nevertheless, one cannot determine specific values
for the first two couplings from the measured
branching ratio, unless $g$ is in the higher part
of its expected range, say $0.6-1.$

The differential $d\Gamma/ds$ distributions in
both cases can be used as additional help for
extracting the values of the coupling constants.
If the value of $g_{\cbsbp}$ will turn out to be
in the ``measurable" range, it will be of great
interest to check the $HM\chi L$ relation
$g_{\cbsbp}/g_{\cdsdp}=M_B/M_D.$

\section{Acknowledgement}

We are in debt to Professor Gad Eilam for
helpful remarks and stimulating discussions.
We also acknowledge discussions with
Drs. Simon Robins, Yoram Rozen and Shlomit Terem
on the feasibility of the relevant detection
experiments. 
Special thanks are extended to the referee, 
who pointed out to us a serious omission in the original version.
The research of P.S. has been
supported in part by the Fund for Promotion of
Research at the Technion.

\newpage

\begin{figure}[b]
\epsfxsize=8 truecm
\epsfysize=4  truecm

\moveright1in\hbox{
\epsffile{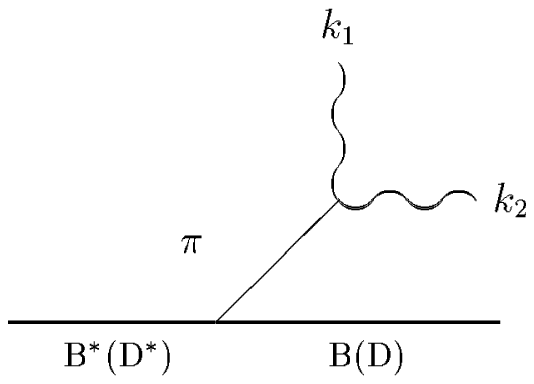}}

\caption{\baselineskip 16pt
The anomaly graph for
$B^*(D^*)\rightarrow B(D)
\gamma\gamma.$
}
\label{fig:1}
\end{figure}

\begin{figure}[b]
\epsfxsize=10 truecm
\epsfysize=4 truecm

\moveright1in\hbox{
\epsffile{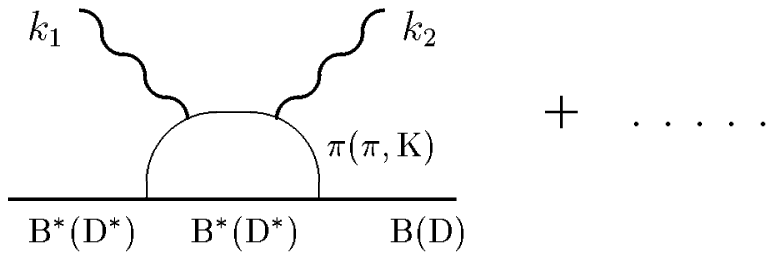}}

\caption{\baselineskip 16pt
Loop graph for $B^*(D^*)\rightarrow B(D)
\gamma\gamma $ which depends only on the strong coupling.
Additional graphs of this kind are discussed in
the text.}
\label{fig:2}
\end{figure}

\begin{figure}[b]
\epsfxsize=14 truecm
\epsfysize=4 truecm

\moveright1in\hbox{
\epsffile{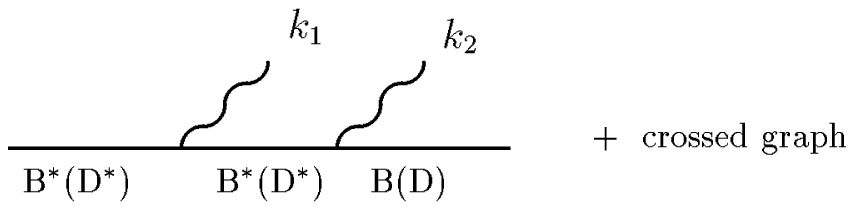}}

\caption{\baselineskip 16pt
Tree graph for $B^*(D^*)\rightarrow B(D)
\gamma\gamma.$}
\label{fig:3}
\end{figure}

\newpage

\begin{figure}[b]
\epsfxsize=14 truecm
\epsfysize=4 truecm

\moveright1in\hbox{
\epsffile{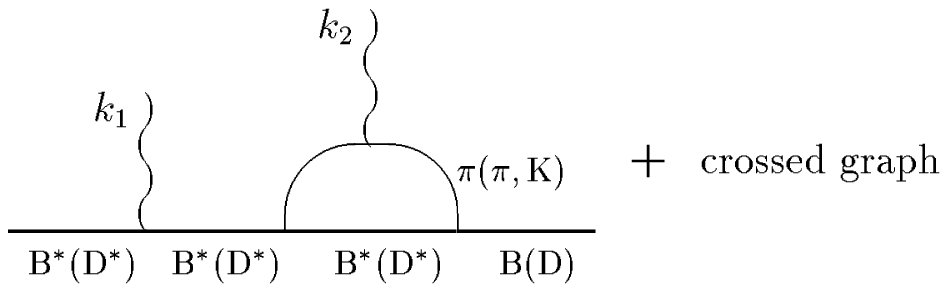}}

\caption
{\baselineskip 16pt
 Loop graph for $B^*(D^*)\rightarrow B(D)
\gamma\gamma$ wich involve both the magnetic
and the strong couplings. Instead of $B^*(D^*)$ in
the first propagator we can have $B(D).$ }
%\label{fig:4}
\end{figure}

\begin{figure}[b]
\epsfxsize=14 truecm
\epsfysize=4 truecm

\moveright1in\hbox{
\epsffile{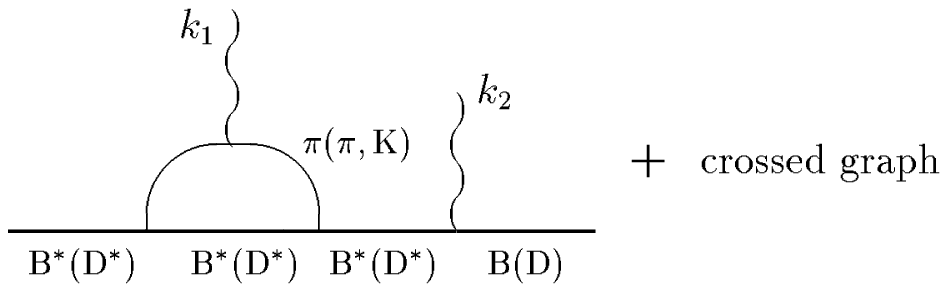}}

\caption{\baselineskip 16pt
Loop graph for $B^*(D^*)\rightarrow B(D)
\gamma\gamma $ wich involve both the magnetic
and the strong couplings. Instead of $B^*(D^*)$ in
the loop we can have $B(D).$}
\label{fig:5}
\end{figure}

\begin{figure}[b]
\epsfxsize=14 truecm
\epsfysize=6 truecm

\moveright1in\hbox{
\epsffile{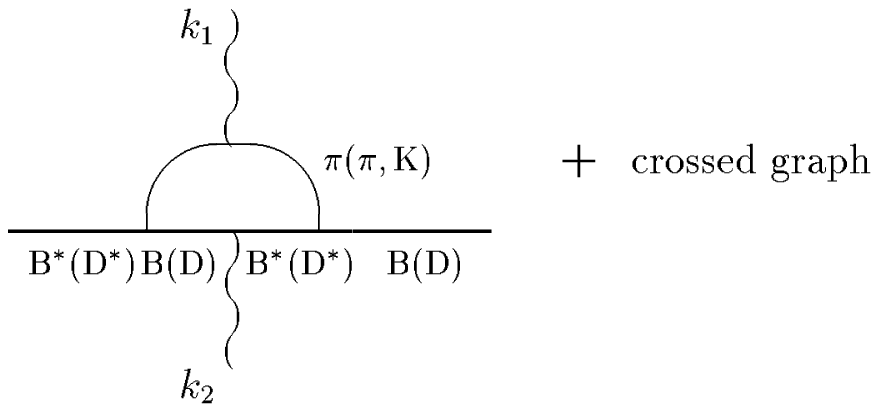}}

\caption{\baselineskip 16pt
Loop graph for $B^*(D^*)\rightarrow B(D)
\gamma\gamma$ wich involve both the magnetic
and the strong couplings.}
\label{fig:6}
\end{figure}

\newpage

\begin{figure}[b]
\epsfxsize=10 truecm
\epsfysize=7  truecm

\moveright1in\hbox{
\epsffile{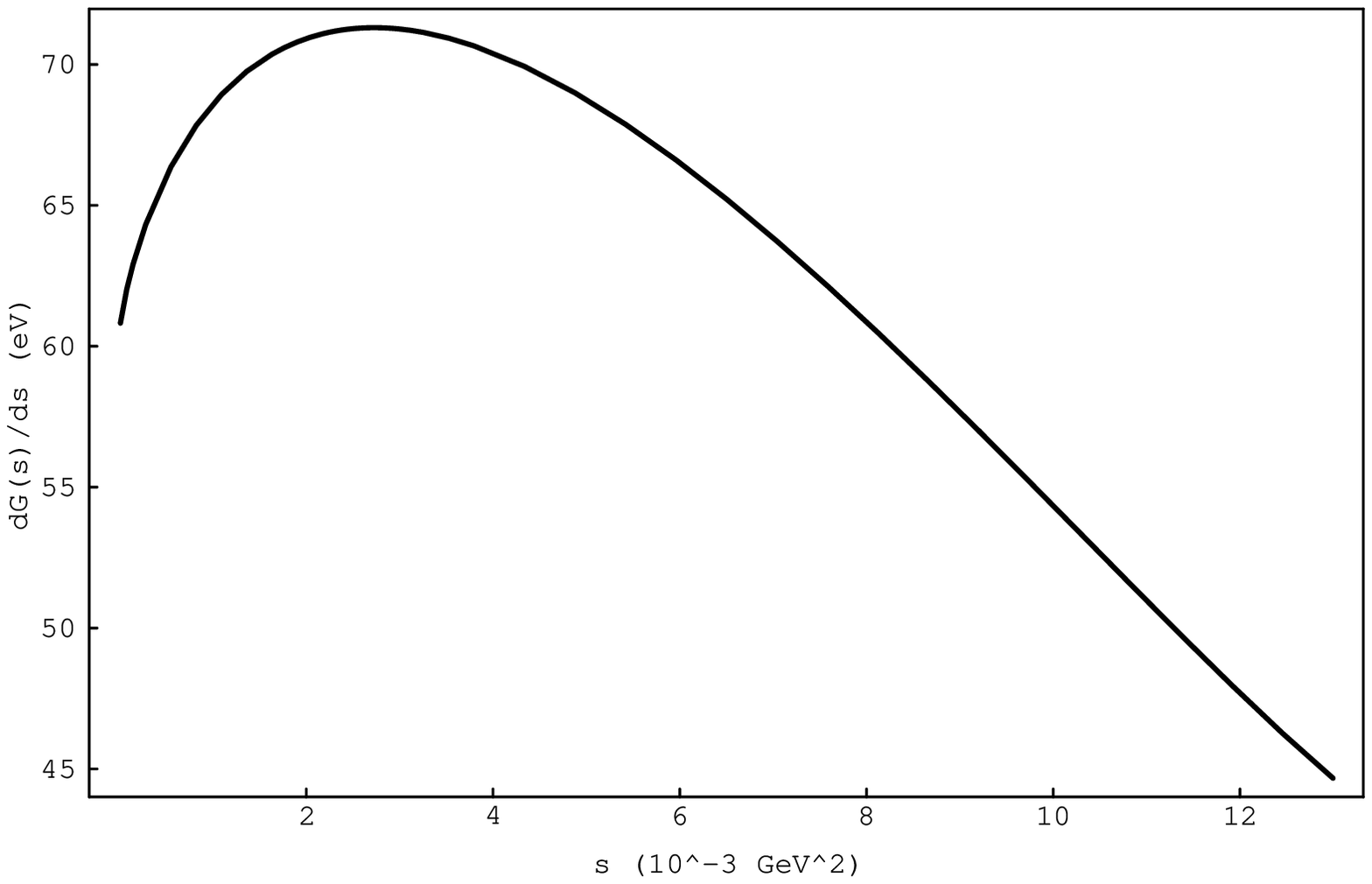}}

%\centerline{\epsffile[40 48 263 213]{glittle1.ps}}
%\centerline{\epsffile[100 203 503 503]{afbpos.ps}}

\caption{\baselineskip 16pt
The  differential decay width $ d\Gamma(\Dsgg)/ds$
(eV)
 as a function of  $s$
with the value  $g=0.7.$ }
     % end tighten
     \end{figure}

\begin{figure}[b]
\epsfxsize=10 truecm
\epsfysize=7  truecm

\moveright1in\hbox{
\epsffile{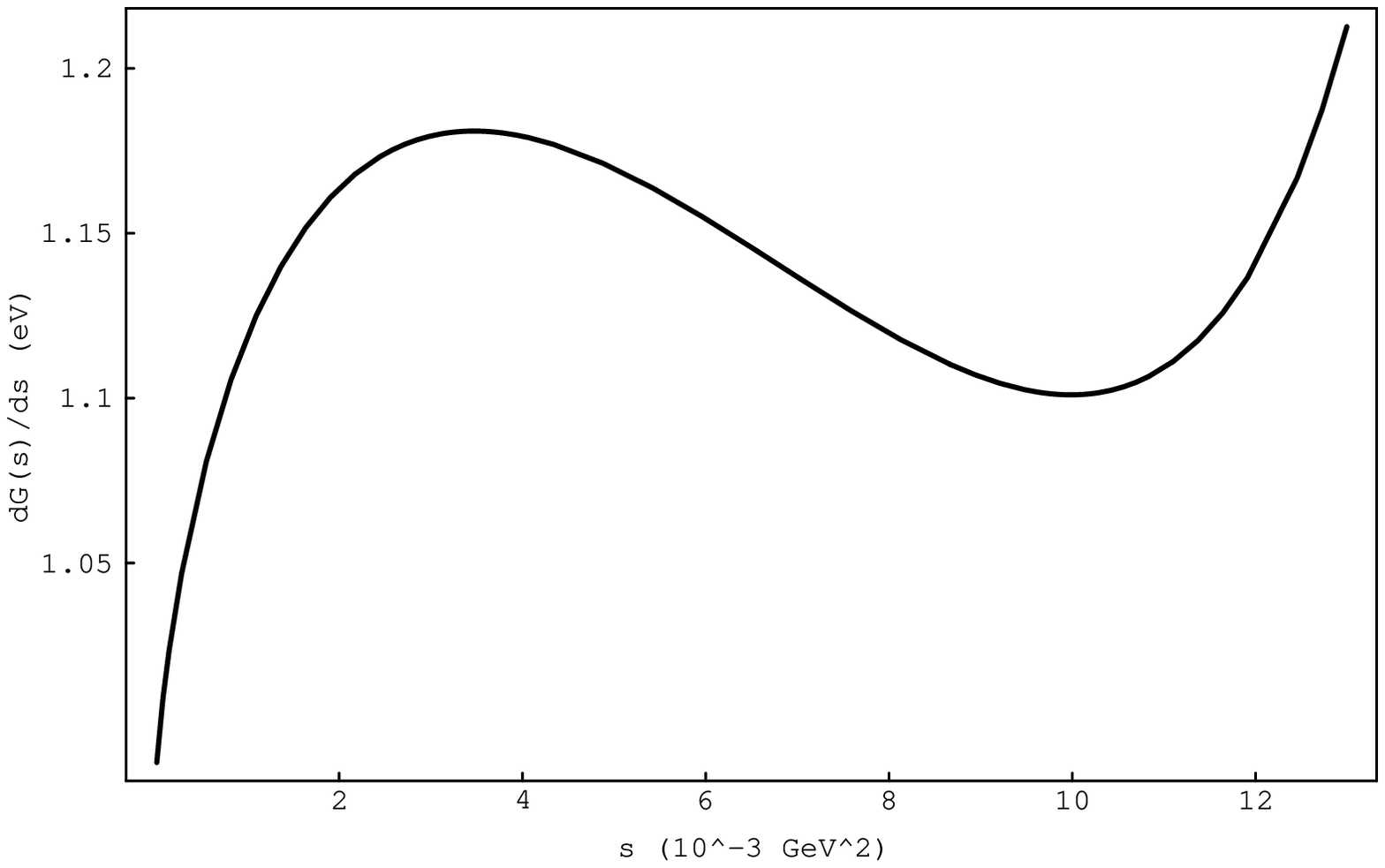}}

%\centerline{\epsffile[40 48 263 213]{glittle1.ps}}
%\centerline{\epsffile[100 203 503 503]{afbpos.ps}}

\caption{\baselineskip 16pt
The  differential decay width $ d\Gamma(\Dsgg)/ds$
(eV)
 as a function of  $s$
with the value  $g=0.25.$ }
     % end tighten
     \end{figure}

\begin{figure}[b]
\epsfxsize=10 truecm
\epsfysize=7  truecm

\moveright1in\hbox{
\epsffile{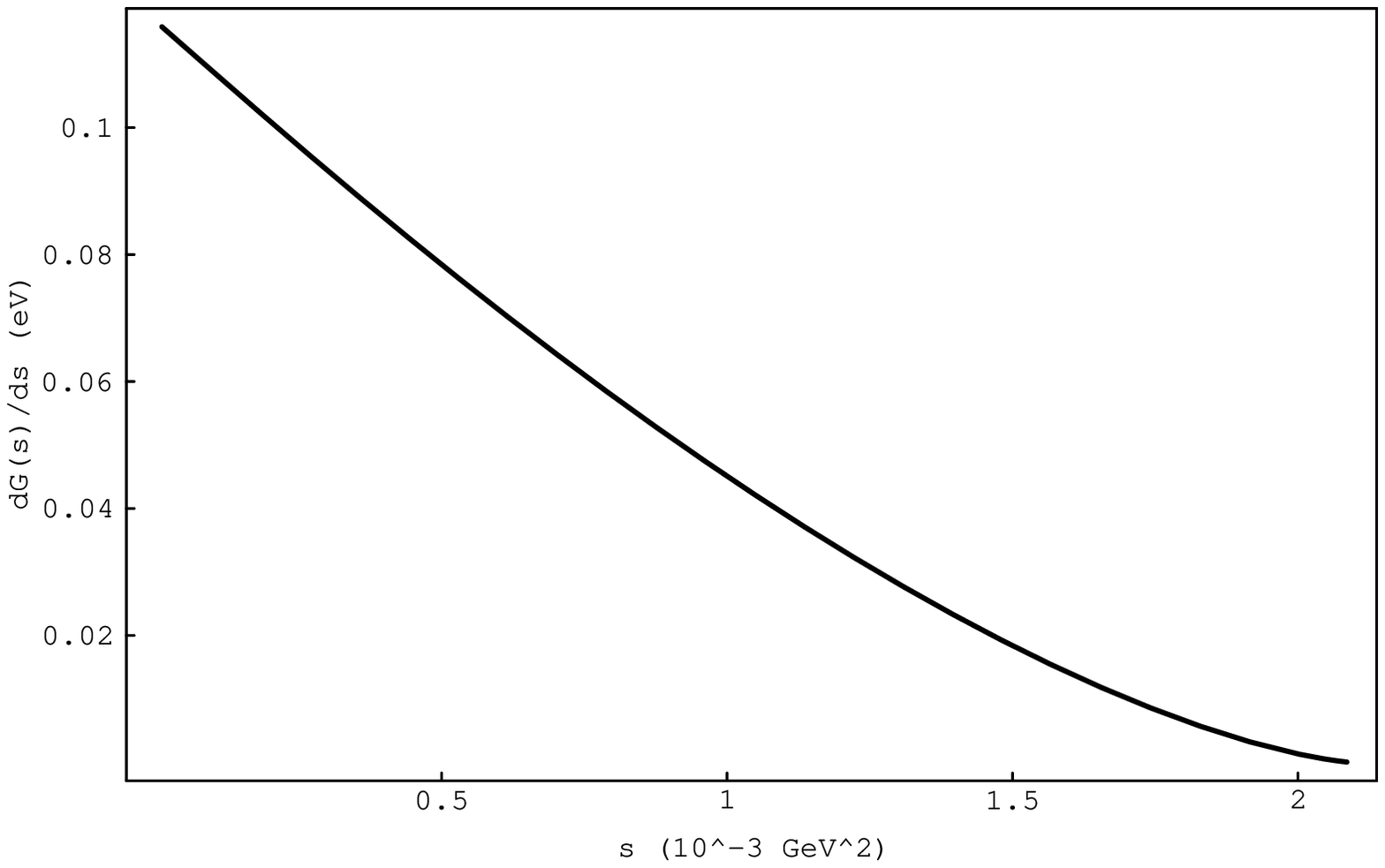}}

%\centerline{\epsffile[40 48 263 213]{glittle1.ps}}
%\centerline{\epsffile[100 203 503 503]{afbpos.ps}}

\caption{\baselineskip 16pt
The  differential decay width $ d\Gamma(\Bsgg)/ds$
(eV)
 as a function of  $s$
with the value  $g=0.5$ and $\mu=2.2.$}
     % end tighten
   \end{figure}

\begin{figure}[b]
\epsfxsize=10 truecm
\epsfysize=7  truecm

\moveright1in\hbox{
\epsffile{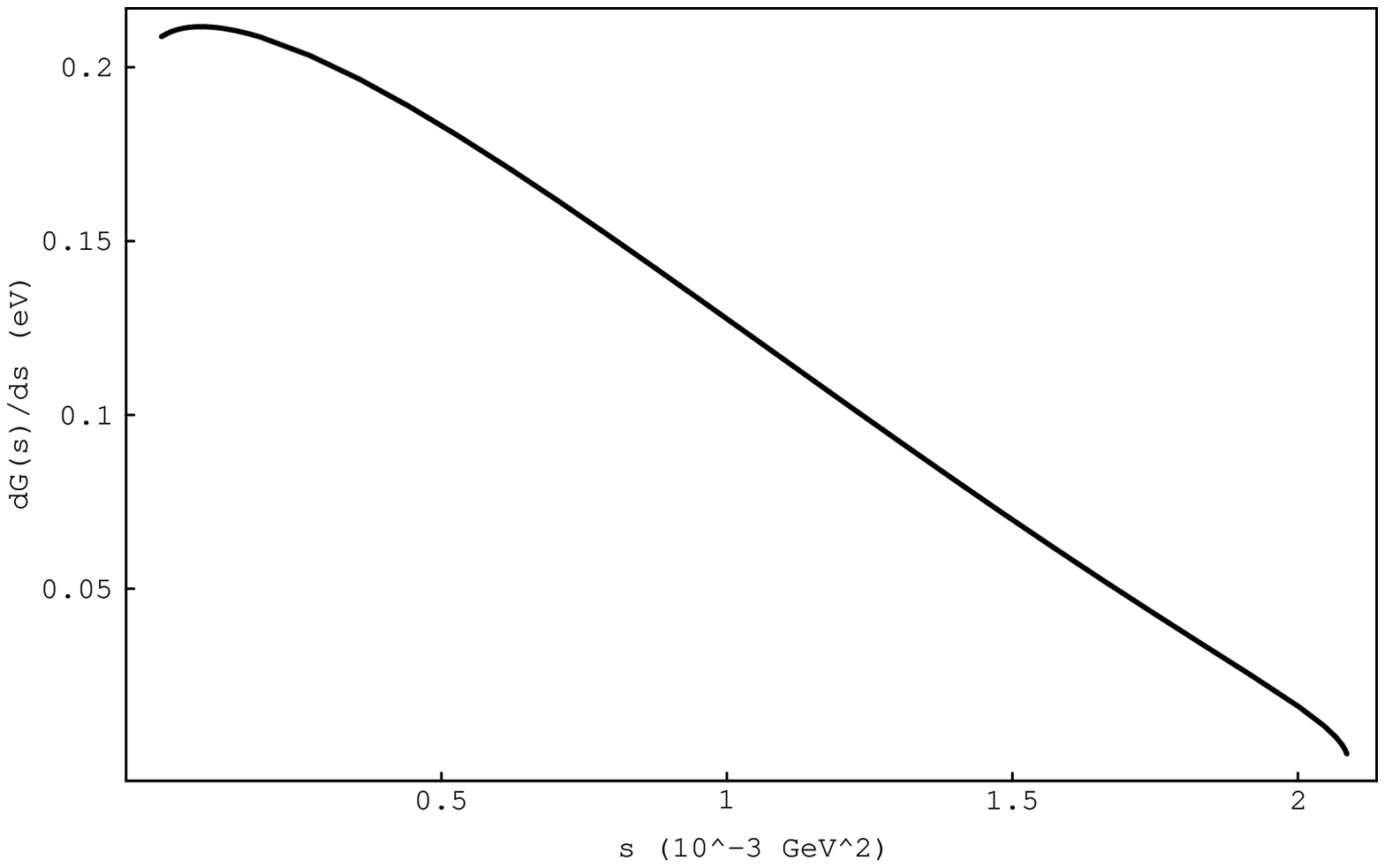}}

%\centerline{\epsffile[40 48 263 213]{gbig1.ps}}
%\centerline{\epsffile[100 203 503 503]{afbpos.ps}}

\caption{\baselineskip 16pt
The differential  decay widths $
d\Gamma(\Bsgg)/ds$ (eV)
 as a function of  $s,$
with the value  $g=0.5$ and $\mu=5.7.$}
     % end tighten
     \end{figure}

     \begin{figure}[b]
\epsfxsize=10 truecm
\epsfysize=7  truecm

\moveright1in\hbox{
\epsffile{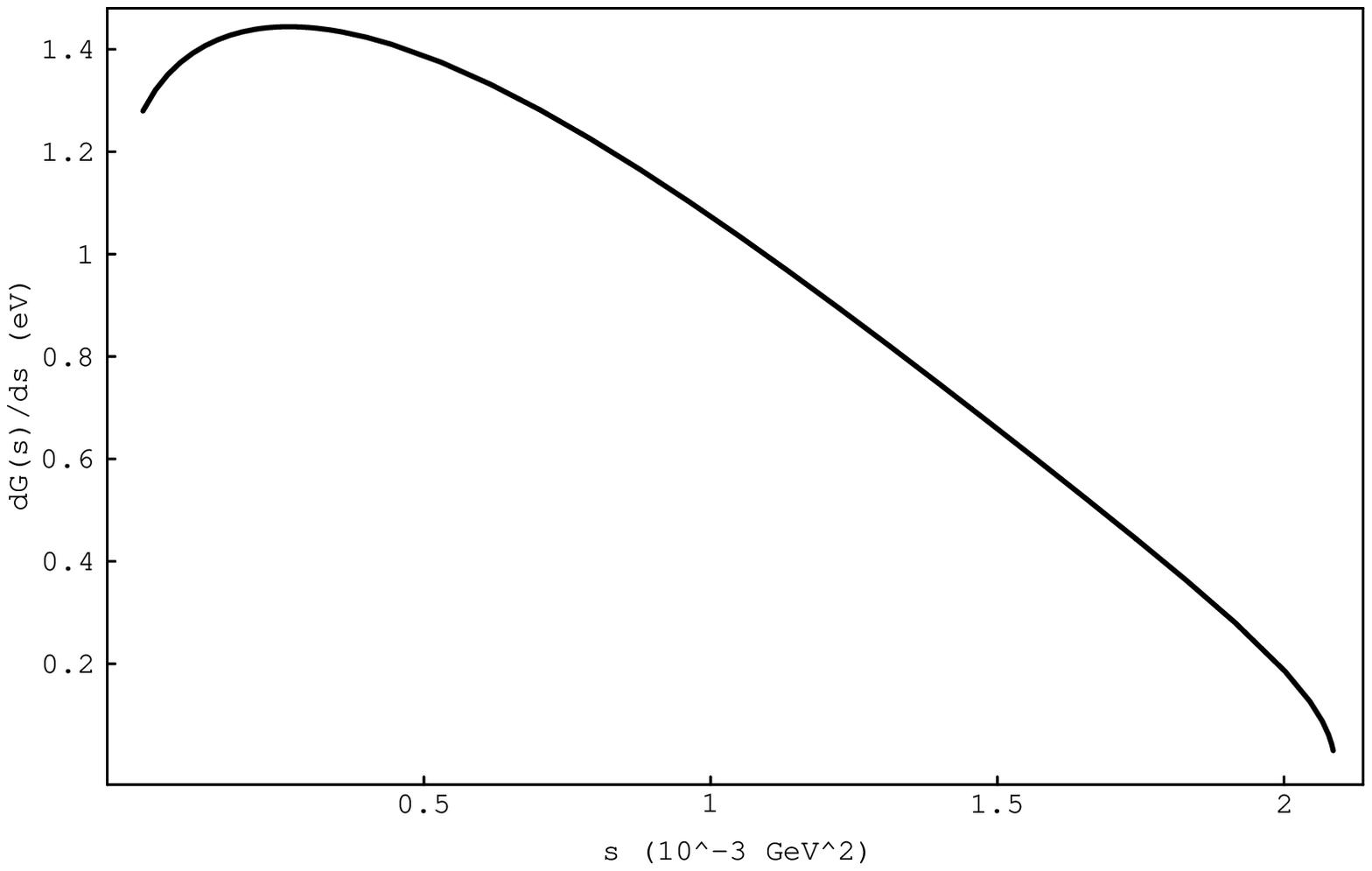}}

%\centerline{\epsffile[40 48 263 213]{glittle1.ps}}
%\centerline{\epsffile[100 203 503 503]{afbpos.ps}}

\caption{\baselineskip 16pt
The differential  decay widths $
d\Gamma(\Bsgg)/ds$ (eV)
 as a function of  $s,$
with the value  $g=0.5$ and $\mu=11.0.$ }
     % end tighten
     \end{figure}

%\begin{figure}[b]
%\epsfxsize=10 truecm
%\epsfysize=7  truecm

%\moveright1in\hbox{
%\epsffile{grafd3.ps}}

%\centerline{\epsffile[40 48 263 213]{GrafD3.ps}}
%\centerline{\epsffile[100 203 503 503]{afbpos.ps}}

%\caption{\baselineskip 16pt
%The differential  decay widths $
%d\Gamma(\Dsgg)/d\sh$ (eV)
% as a function of  $\sh,$
%with the value  $g=0.38.$
%The anomaly contribution (solid),
%the loop one (dashed) and the sum of the two (dash-dotted).}
     % end tighten
%     \label{figd}
%\end{figure}

%%%%%%%%%%%%%%%%%%%%%%%%%%%%%%%%%%%%%%%%%%%%%%%%%%%%%%%%

\newpage
%%%%%%%%%%% TABLE 2 %%%%%%%%%%%%%%%%%%%%%%%%%%%%%%
%
\begin{table}[h]
\begin{center}
 {\tighten
\caption{\baselineskip 16pt
Predictions for the various $D^*\rightarrow D
\gamma\gamma$  decay  for various values of
$g$ (eqs.(5.9) and (5.12))}
\vspace{0.3cm}
\begin{tabular}{ c  c  c}
$g$ & ${\rm BR}(D^*)(\bar{\mu}=6.6 g,\bar{\mu}_+=1.7g)$  &$  {\rm
BR}(D^*)(\bar{\mu}=-6.6g,\bar{\mu}_+=-1.7g)$
\\
  \hline
\noalign {\vspace{1truemm} }
$ g=0.25
         $&$ 1.7\times 10^{-6}
 $&$ 1.7 \times 10^{-6} $

\\$ g=0.38
         $&$  3.9 \times 10^{-6}  $&$ 3.7\times 10^{-6} $
\\ $ g=0.5
         $&$    6.9 \times10^{-6} $&$ 6.3\times 10^{-6}$
\\ $g=0.7  $&$ 1.4 \times10^{-5}
         $&$ 1.3 \times 10^{-5} $ \\
$g=1 $&$ 3.3\times 10^{-5} $&$ 2.9\times
10^{-5}$\\
\noalign {\vspace{1truemm} }
\end{tabular}
 }     % end tighten
\end{center}
\end{table}
%

%%%%%%%%%%%%%%%%%%%%%%%%%%%%%%%%%%%%%%%%%%%%%%%%%%%%%%%%

\begin{table}[h]
\begin{center}
 {\tighten
\caption{\baselineskip 16pt
Predictions for the various $B^*\rightarrow B
\gamma\gamma$  decay  different values of
$g$ and $\mu$}
\label{table2}
\vspace{0.3cm}
\begin{tabular}{ c  c  c  c}
$g$ & ${\rm BR}(B^*)(\bar{\mu}=2.2)$  &$  {\rm
BR}(B^*)(\bar{\mu}=11.0)$ &${\rm
BR}(B^*)(\bar{\mu}=-2.2)$
\\
  \hline
\noalign {\vspace{1truemm} }
$ g=0.25
         $&$ 3.1\times 10{^-7}
 $&$ 1.9 \times 10^{-6} $&$2.4\times 10^{-7}$
\\$ g=0.38
         $&$  1.3 \times 10^{-6}
         $&$ 2.2\times 10^{-6} $&$ 9.4 \times 10^{-7}$
\\ $ g=0.5
         $&$    3.7 \times10^{-6} $&$ 2.9\times 10^{-6}$
         &$2.7\times 10^{-6}$
\\ $g=0.7  $&$ 1.4 \times10^{-5}
         $&$ 5.5 \times 10^{-6} $&$9.2\times 10^{-6}$ \\
$g=1 $&$ 4.8\times 10^{-5} $&$ 9.0\times
10^{-6}$&$3.8\times  10^{-5}$\\
\noalign {\vspace{1truemm} }
\end{tabular}
 }     % end tighten
\end{center}
\end{table}
%

%%%%%%%%%%%%%%%%%%%%%%%%%%%%%%%%%%%%%%%%%%%%%%%%%%%%%%%%

\begin{thebibliography}{99}

\bibitem{cusb}
CUSB Collab., K. Han et al., \prl{55}, 36 (1985).


\bibitem{cusb2}
CUSB-II Collab., J. Lee-Franzini et al., \prl{65}, 2947
(1990).


\bibitem{opal}
ALEPH Collab., Z. Phys.  C 69, 393 (1996); DELPHI
Collab., Z. Phys. C 68, 353 (1995); L3 Collab.,
\plb{345}, 589 (1995); OPAL Collab., Z. Phys. C 74, 413 (1997).

\bibitem{PDG}
Particle Data Group, C. Caso et al, Europ. Phys.
Journal C3, 1 (1998).

\bibitem{Eichten}
E. Eichten et al., \prd{21}, 203 (1980).

\bibitem{Godfrey}
S. Godfrey and N. Isgur, \prd{32} , 189 (1985).

\bibitem{Ivanov}
M.A. Ivanov and Yu. M. Valit, Z. Phys. C 67 , 633
(1995).

\bibitem{Singer}
P. Singer and G. A. Miller, \prd{34}, 825 (1989).

\bibitem{Cho}
P. Cho and H. Georgi, \plb{296}, 408 (1992).

\bibitem{Amundson}
J.F.Amundson et al, \plb{296}, 415 (1992).

\bibitem{Colangelo}
P. Colangelo, F. De Fazio and G. Nardulli,
\plb{316},  555 (1993).

\bibitem{Bank}
N. Barik and P.C. Dash, \prd{49},299 (1994).

\bibitem{DeFazio}
P. Colangelo, F. De Fazio, and G. Nardulli,
\plb{334}, 175 (1994).

\bibitem{Dosch}
H.G. Dosch and S. Narison, \plb{368}, 163 (1996).

\bibitem{Aliev}
T. M. Aliev et al, \prd{54}, 857 (1996).

\bibitem{Peruzzi}
 I. Peruzzi et al., \prl{37},  569 (1976);
 Mark I Collab., G.Goldhaber et al, \plb{69}, 503 (1977);
G. J. Feldman et al.,
\prl{38},  1313 (1977).

\bibitem{JADE}
JADE Collab., W. Bartelt et al,  Phys. Lett. {\bf 161 B}, 197
(1985).

\bibitem{HRS}
HRS Collab., E.H. Low et al, \plb{183}, 232
(1987), S. Abachi et al.,
\plb{212}, 533 (1988).

\bibitem{CLEO}
CLEO Collab., F. Butler et al. \prl{69}, 2041 (1992).

\bibitem{ARGUS}
ARGUS Collab., H. Albrecht et al, Z. Phys. C 66, 63 (1995).

\bibitem{CLEO1}
CLEO Collab., J. Bartelt et al, \prl{80}, 3919  (1998).

\bibitem{ACCMOR}
ACCMOR Collab., S. Barlag et al
\plb{278}, 480 (1992).

\bibitem{Thens}
R.L. Thews and A.N. Kamal,\prd{32}, 810 (1985); E.
Sucipto and R.L. Thews, \prd{36}, 2074 (1987).

\bibitem{Pham}
T.N. Pham, \prd{25}, 2955 (1982).

\bibitem{Donnel}
P. J. O'Donnel and Q. P. Xu,
\plb{336}, 113 (1994).

\bibitem{Miller}
G. A. Miller and P. Singer, \prd{37}, 2564 (1988).

\bibitem{Belyaev}
V.M. Belyaev, V.M. Braun, A. Khodjamirian and R.
Ruckl, \prd{51}, 6177 (1995).

\bibitem{Troytskaya}
A.N. Ivanov and N.I. Troytskaya, \plb{345}, 175
(1995).
\bibitem{Kamal}
 A. N. Kamal and Q. P. Xu,
\plb{284}, 421 (1992).

\bibitem{Stewart}
I. Stewart, \npb{529}, 62 (1998).

\bibitem{Wise}
M.B. Wise, \prd{45}, R2188 (1992).

\bibitem{Burdmann}
G. Burdmann and J. Donoghue, \plb{280}, 287
(1992).

\bibitem{Yan}
T.M. Yan et al, \prd{46}, 1148 (1992).

\bibitem{Casalbuoni}
R. Casalbuoni et al, Phys. Rept. {\bf 281}, 145
(1997).

\bibitem{Cheng2}
H.-Y. Cheng et al, \prd{47}, 1030 (1993); {\bf
55},5851 (E) (1997).

\bibitem{Cheng}
H.-Y. Cheng et al, \prd{49}, 2490 (1994); {\bf
55},5851 (E) (1997).

\bibitem{Boyd}
C.G. Boyd and B. Grinstein, \npb{442}, 205 (1995).

\bibitem{Leibovich}
A.K. Leibovich, A.V. Manohar and M.B. Wise,
\plb{358}, 347 (1995);376, 332(E) (1996).


\bibitem{Cola}
P. Colangelo et al, \plb{339}, 151 (1994).


\bibitem{Manohar}
A.V. Manohar and H. Georgi, \npb{234}, 189 (1984).

\bibitem{Isgur}
N. Isgur and M.B. Wise, \prd{41}, 151 (1990).

\bibitem{Nussinov}
S. Nussinov and W. Wetzel, \prd{36}, 130 (1987).


\bibitem{UKQCD}
UKQCD Collab., G. M. de Divitiis et al,
hep-lat/9807032.

\bibitem{Chao}
N.-G. Chen and K.-T. Chao, \plb{345}, 67 (1995).

\bibitem{Dafne}
D.Guetta and P.Singer (to be published).

\end{thebibliography}
\end{document}